# Nature of the growth of plasma electrolyte oxidation coating on Aluminum

Lujun Zhu, Xiaoxing Ke, Jingwei Li, Yuefei Zhang, Zhenxin Chen, Zhenhua Zhang, Yue Lu, and Manling Sui *

Institute of Microstructure and Property of Advanced Materials, Beijing University of Technology, Beijing 100124, China

(December 7, 2017)

**Abstract**

The localized dielectric breakdown had been considered as the driving force for growth of plasma electrolytic oxide (PEO) coatings for several decades. However, the TEM study here reveals the dielectric breakdown behavior has little contribution for coating thickening. The presented evidences show the nature of PEO coating growth in all three consecutive stages I-III is ionic migration behavior inside the amorphous alumina layer (AAL) at the coating/matrix interface. The evolution of morphological characterizations in the PEO process is attributed to the interfacial reactions of AAL/alkaline (stage I), AAL/discharge-on-surface (stage II) and AAL/discharge-in-pore (stage III).

Keyword：Aluminum, SEM, TEM, anodic films, oxide coatings

## 1. Introduction

Plasma electrolytic oxide (PEO), also known as anodic spark deposition (ASD) [1] or micro arc oxidation (MAO) [2], is an surface engineering technology of forming a thick ceramic coating on valve alloys (e.g., Al, Ti, Mg, Nb) by means of a micro discharge. Based on process controls, the PEO coating could contain various ceramic components [3] and incorporation particles [4, 5]. Considered as an eco-friendly and simple technical process, the PEO technology has been developed as one of the most promising and versatile methods to improve the wear, friction, corrosion, thermal and other desirable properties of materials [6-9].

In nature, the formation of PEO coating on Al is a complex process combining with oxidation of metal, deposition of electrolyte and discharge in liquid. According to traditional theory, the whole process of the PEO coating was thought to contain a conventional anodic oxidation stage and a follow-up micro-discharge stage [10-12], and the cut-off point of the two stages was the first surface discharge (also called sparking initiation). In the anodic oxidation stage, an amorphous-porous coating formed and grew upto several hundred nanometers via ionic migration process [1,10]. And then, the dielectric-amorphous coating would be breakdown when the voltage exceeded the threshold value, and the discharge spark occurred [13, 14]. In the latter micro-discharge stage, the PEO coating grew continuously upto several ten micrometers via a violent ejection process, i.e. the localized dielectric breakdown mechanism [8, 15-17]. It was thought to include three steps: (i) the dielectric coating is broken down and an instantaneous temperature rise emerges at the discharge zone; (ii) the localized aluminum is molten into the channel and the aluminum atoms are oxidized; (iii) when the discharge is quenched, the aluminum oxides, mainly alumina, in the channel are cooled down and deposited onto the channel walls, resulting in the coating growth. This dielectric breakdown mechanism could be described as "breakdown-melt-ejection-deposition" in brief and was established based on the high concentration of energy induced by a localized electron avalanche effect.

In contrast to what is described by the traditional viewpoint, the PEO process has been found to contain at least three consecutive stages in many ways, such as electrical characteristics [11], structure morphology [18], acoustic emission [19] and optical emission [20]. Moreover, based on the calculation results of the optical emission spectra collected from the discharge during the PEO process, the electron density was determined to be $10^{15}$-$10^{17}$ cm$^{-3}$ in magnitude [21], which can only be treated as a soft gas plasma [22] rather than the dielectric breakdown [23, 24]. Our recent microstructure study on the micro-discharge stage of the PEO process has revealed that the PEO coating growth is not a violent ejection of molten materials, but a gentle growing process via an ionic migration mechanism [25]. The cross-sectional microstructure characterization showed that the discharges ignited in the gas-filled channels, which passed through both the outer and the internal coating layer, and the amorphous alumina layer (AAL) as the innermost coating layer was thickened by transporting $Al^{3+}$ and $O^{2-}$ ions under the loading voltage, meanwhile the thickened internal AAL extruded the PEO coating to grow outward [25]. However, up to now, there is no systematic microstructure study to reveal the various formation mechanisms in different stages of the whole PEO process, especially for the initial stages, which has been linked directly to the corrosion [26] and photocatalytic properties [9].

The present work was specially aimed at a systematic investigation on the composition, morphology and structure characteristics of the initial stages in the PEO process on Al, by employing scanning electron microscopy (SEM), and transmission electron microscopy (TEM). Combining with our previous work on the micro-discharge stage [25], the formation and growth mechanisms were proposed to describe the different stages in the whole process of the PEO coating on Al.

## 2. Experimental

Aluminum (wt%≥99.7) samples with the dimensions of 10 mm ×25 mm ×0.3 mm were used in this experiment, which were ultrasonically cleaned in ethanol and acetone for 5 min respectively, and then dried under a nitrogen flux. The PEO treatment was carried out in a 15 L stainless steel container, equipped with a stirrer and water-cooling to keep the electrolyte temperature below 27 ˚C. In the experiment, the sample and the wall of the container were used as two electrodes respectively, and the container was filled with an electrolyte solution (containing 12.5 g $L^{-1}$ $Na_2SiO_3$ and 5 g $L^{-1}$ KOH with a pH value of ~13). The coating deposition was carried out by a unipolar pulse at a constant mean current density of 0.1 A $cm^{-2}$, with a duty cycle of 40% and frequency of 100 Hz, using a 15 kW pulsed power supply (Pulsetech Electrical Co., Ltd., Chengdu,

China). The current change was manually recorded as displayed on the instrument. The cell voltages were regarded as the peak values of the potential pulses, measured by an Agilent 33410A digital multimeter with a counting interval of 0.006 s.

After the coating fabrication, the samples were observed by an environment-SEM (ESEM) directly and then coated with a 10-nm thick Au film to acquire the high resolution SEM images. The cross-sectional TEM samples for the coating layers on the Al substrate were prepared using a focused ion beam (FIB, FEI Helios Nanolab 600i system). A final cleaning for the sample surface was performed using 2 keV Ga+ ions with a small beam current to minimize the ion beam damage on the coating layers.

The SEM observations in both the back-scattered electron (BSE) model and the secondary electron (SE) model were made by a FEI Quanta 250 ESEM at 60 Pa. The spherical aberration corrected (Cs-corrected) high angle annular dark field scanning transmission electron microscopy (HAADF-STEM) and the energy-dispersive X-ray (EDX) mapping experiments were performed using a FEI Titan G2 microscope, equipped with a Super-X detector, operated at 300 keV.

**3.Result and discussions**

3.1. Evolution of electrical characteristics in the PEO process

The evolution of the principal electrical characteristics (voltage vs. time) at the first 30 minutes of the PEO process is shown in Fig. 1, which exhibits three typical consecutive stages: the stage I at the initial ~15 s with a rapidly increasing voltage, the stage II at 15 s – 9 min with a slow increasing, and the stage III with a nearly stable voltage after ~ 9 min. To focus on the initial 20 s of the PEO formation process, the voltage vs. time curve is exhibited by the red line in Fig. 2a. The inset images of Fig. 2a are the sample surface morphology at the moment of 4 s, 8 s, 12 s and 18 s, respectively. The blue line-symbol plot in Fig. 2a shows the increase of the mean current density, which took about 5 s to reach the set point of 0.1 A·cm$^{-2}$. To avoid the influence of the unsteady current density at the initial 5 s, a R-value, the ratio of the voltage and the current density, was used here to qualitatively describe the voltage change at the initial 20 s of the PEO process. As shown in the R-value vs. time curve (Fig. 2b), the first 20 s of the PEO process can be divided into four segments:

(i). In the beginning of 1s, the R-value is relatively low and decreases from 978 to 474 $\Omega\cdot cm^2$, which implies the dissolution [12] or transformation [27] of the pre-existed passive surface layer when the voltage is over 10 V.

(ii). In 1-11 s, the R-value increases linearly with a rate of ~184 $\Omega\cdot cm^2/s$, corresponding to the stage I, i.e. the anodic oxidation process along with the luminescence at the sample surface (shown as the insets of 4 s and 8 s in Fig. 2a).

(iii). In 11-15 s, the R-value increases at a rate of ~385 $\Omega\cdot cm^2/s$, corresponding to a breakdown stage (BD-stage) with the initiated discharge spark (shown as the inset of 12 s in Fig. 2a).

(iv). Beyond 15s, the increasing rate of R-value is only ~23 $\Omega\cdot cm^2/s$, corresponding to the stage II (Fig. 1) with the enhanced discharge spark (shown as the inset of 18 s in Fig. 2a).

It should be mentioned that the increasing rate of R-value will decrease to ~0.06 $\Omega\cdot cm^2/s$ in the stage III (beyond 9 min, Fig. 1). In the aspect of electrical characteristics, the whole PEO process contains at least three consecutive stages.

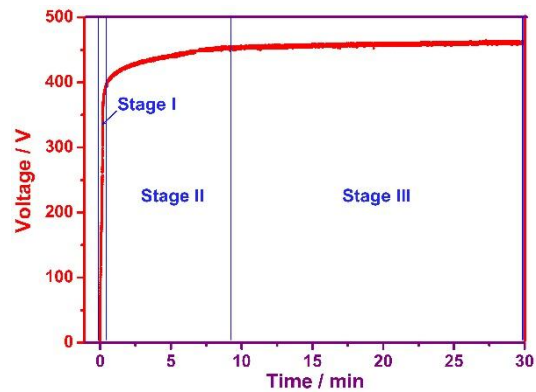

Fig. 1. The curve of peak voltage vs. time at the first 30 min of the PEO process.

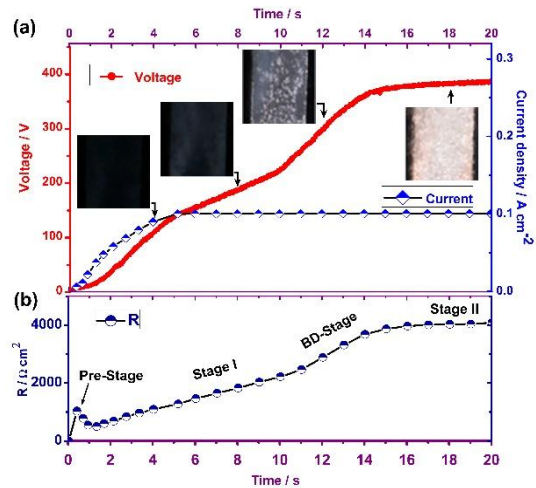

Fig. 2. The curves of (a) voltage (red line) vs. time, current (blue line-symbol plot) vs. time and (b) R-value vs. time at the initial 20 s of the PEO process. The insets in Fig. 2a are the sample surface morphology at the time of 4 s, 8 s, 12 s and 18 s, respectively. The R-value refers to the ratio of the voltage and the current density.

3.2. Structure of the PEO coating in the stage I.

The SEM and TEM observations are applied to characterize the structural features in three stages of the PEO coating. Figs. 3a-c are the surface micrographs of the PEO coating anodized for 0 s, 4 s and 8 s in the stage I, respectively. The stage I of the PEO coating is widely considered as the ionic migration process, in which the $Al^{3+}$ and $O^{2-}$ inside amorphous alumina coating are transported by electric migration of high field strength ($E$), and the amorphous alumina layer (AAL) is formed with the parallel-sided pore arrays inside [25, 28]. In Fig. 3c, it is easy to distinguish the dense pores with diameter of ~100 nm distributing uniformly at the 8 s-sample surface.

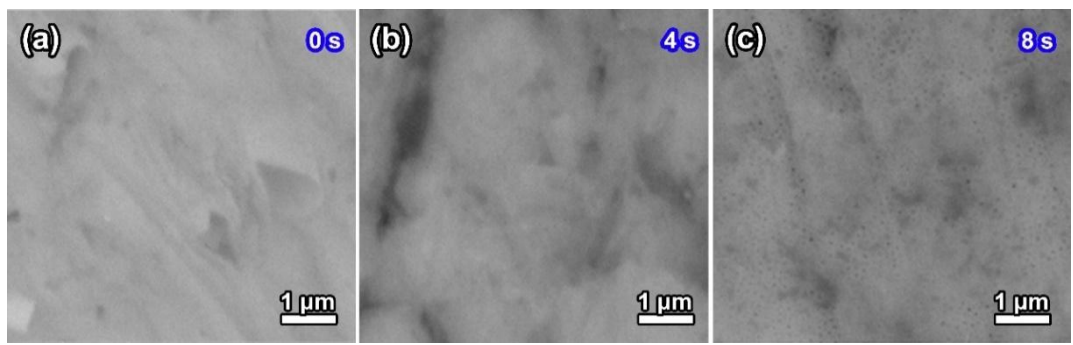

Fig. 3. SEM (BSE model) surface micrographs of the PEO coating anodized for (a) 0 s, (b) 4 s, and (c) 8 s.

To reveal these pores, the cross-sectional TEM specimens were prepared by perpendicularly cutting the PEO coating surface using FIB. As shown in Fig. 4a, the cross-sectional TEM micrograph of the 4 s-sample reveals the Al matrix, the formed AAL (the PEO coating layer) and the Pt protection layer (deposited when sample fabricated by FIB) from bottom to up. Texture was noticed in the AAL layer with a thickness of 500 nm. Fig. 4b is the enlarged image of Fig. 4a, which exhibits two parallel-sided pore channels and the concave Al/AAL boundary. The pore with the largest diameter is selected as the representative to show the AAL geometry of the PEO coating anodized for 4 s. As shown in Fig. 4c, the distance between the pore bottom and the concave Al/AAL boundary is ~147 nm and the measured voltage is ~150 V (Fig. 2a), then the $E$ applied on the AAL is ~1 V·nm$^{-1}$, which is the typical $E$ value required for the AAL growing via the ionic migration process [25, 28, 29]. Fig. 4d is the cross-sectional TEM micrograph of the PEO coating anodized for 8 s, which shows the AAL grows into the thickness of ~1000 nm. Fig. 4e and Fig. 4f are the corresponding enlarged images and the schematic AAL geometry, respectively. Compared with the 4 s-sample, the pore diameter in the 8 s-sample increases from 15 nm to 70 nm, but the outer diameter of the pore channel remains unchanged. Therefore, the growth of the PEO coating in the stage I is a uniformly thickening process of the ALL via the ionic migration mechanism.

It should be noticed that the distance between the pore bottom and the concave Al/AAL boundary increases to ~350 nm when anodized for 8 s (Fig. 4f), and the measured voltage is ~190 V, therefore the $E$ across the AAL drops to only ~0.54 V·nm$^{-1}$. According to the ionic migration mechanism, when keeping the current density ($I$) constant, the variation of $E$ is proportional to the temperature ($T$) change for the AAL with a slope [30] of:

$$\Delta E/\Delta T = \frac{k_B}{qa} \cdot \ln\frac{I}{I_0},$$

where $k_B$ is the Boltzmann constant, $q$ is the valence charge, $a$ is the jump distance and $I_0$ is a pre-exponential factor for the ionic conductivity in a solid electrolyte [31]. Corresponding to the $E$ decreasing from 1 V·nm$^{-1}$ to 0.54 V·nm$^{-1}$, the $T$ would increase from 23 $^o$C to 250 $^o$C. It means that at the end of the stage I, the temperature of the AAL was estimated to reach the crystallization temperature of the amorphous alumina [32], which would promote the dissolution of the coating surface and then lead to the breakdown of the coating [33].

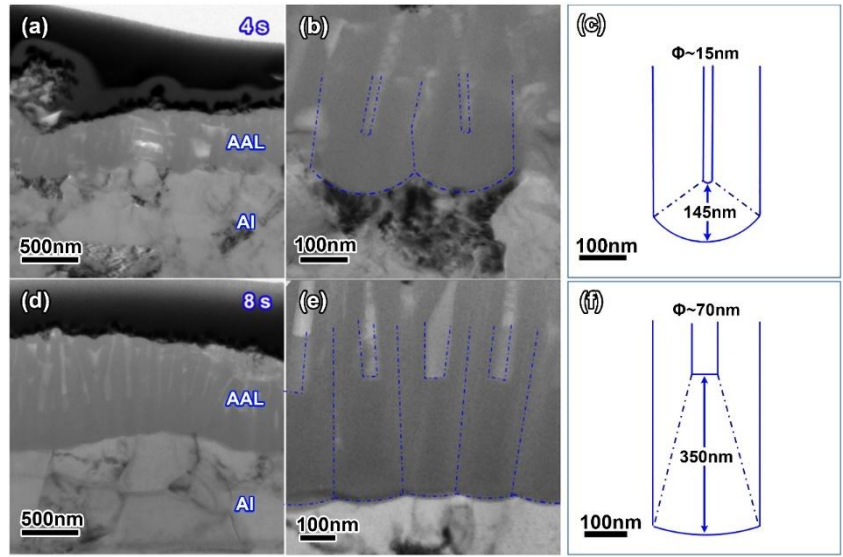

Fig. 4. (a) TEM cross-sectional micrograph, (b) the enlarged image and (c) the AAL geometry of the PEO coating anodized for 4 s. (d) TEM cross-sectional micrograph, (e) the enlarged image and (f) the AAL geometry of the PEO coating anodized for 8 s.

3.3. Structure of the PEO coating in the breakdown stage.

When anodized for 12 s, the PEO process developed into the BD-stage. The breakdown appeared suddenly at the surface of the PEO coating with the initiated discharge spark (the inset of 12 s in Fig. 2a). A large number of breakdown pores with diameter of >1 μm emerged on the sample surface, as shown in the SEM (BSE model) micrograph (Fig. 5a) and the high resolution SEM micrograph (Fig. 5b). Meanwhile, the pores of the ionic migration still existed, but became rare and discrete (Fig. 5b). Fig. 5c is the cross-sectional TEM micrograph of the ionic migration pores with the concave boundary, which reveals the distance between the pore bottom and the concave boundary is ~400 nm. Accordingly, the *E* and *T* inside the AAL are estimated as ~0.75 V·nm$^{-1}$ and 147 $^{o}$C, respectively. As the outcome of the localized discharges in the PEO process, the decrease of temperature implies that the abundant micron-sized breakdown pores provide an approach to dissipate the heat of the PEO coating into the liquid electrolyte.

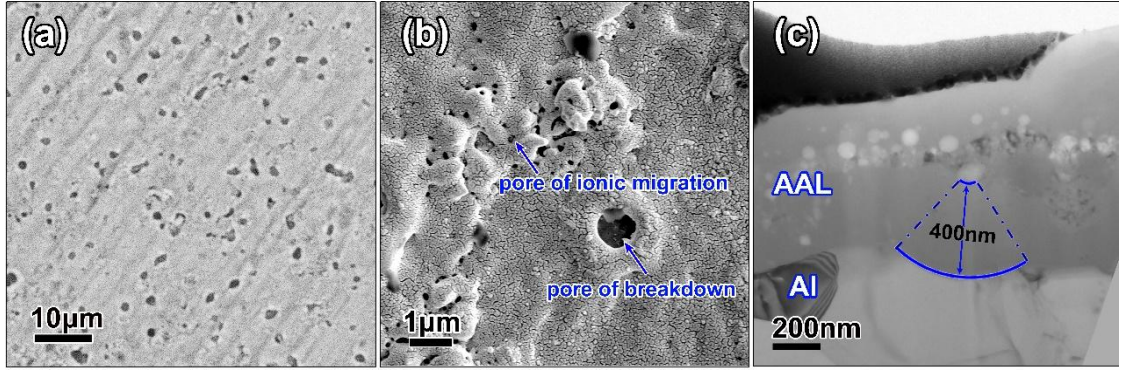

Fig. 5. (a) SEM (BSE model) and (b) high resolution SEM (SE model) micrographs of the coating surface, (c) TEM cross-sectional micrograph of the PEO coating anodized for 12 s.

Fig. 6a is the cross-sectional HAADF-STEM image of the breakdown pore in Fig. 5b. The breakdown pore is a concave pit with the depth of ~500 nm, which is the same as the thickness of the discharged oxide coating, as indicated by the blue line in the remaining part of the coating in Fig. 6a. The oxide layer at the bottom of the pit (indicated by the yellow line in Fig. 6a) is a little thinner than the remaining part of the coating. Fig. 6b shows the EDS mapping of the elements for O, Al and Si by red, blue and green, respectively. The electrolyte Si at the breakdown-pore bottom is distributed more closer to the Al matrix. Based on the theoretical prediction of the breakdown model (as illustrated schematically in Fig. 7), the breakdown of oxide coating contains two processes: (i) The breakdown occurs randomly at some point of the coating surface (Fig. 7a). At the moment, the electron avalanche will cause a plasma evaporation of the oxide, which results in the loss of localized coating and a high thermal stress around the breakdown point [13] (Fig. 7b). The cross-sectional image of the breakdown pore in Fig. 6a shows clearly the loss of oxide coating and the extruding deformation by thermal stress at both sidewalls. (ii) After that, the exposed metal at the pit surface quickly reacts with the electrolyte and then the anodic oxide film is rapidly formed [34] owing to the extremely high current density at the bare spot (Fig. 7c). As shown in Fig.

6b indicated by the arrows, the element mapping of the breakdown pore demonstrates that the Si, which can only come from the electrolyte, appeared more close to the Al matrix due to the high current density at the breakdown-pore bottom, compared to the neighboring area without breakdown. Finally, the anodic oxide coating at the breakdown-pore bottom grows into the similar thickness with the other part of the coating, and the deep pit remains on the surface.

In contrast to the traditional "breakdown-melt-ejection-deposition" mechanism on the PEO coating growth [15-17], the above evidence clearly shows the high concentration of energy induced by a localized breakdown makes local coating loss and creates the breakdown pit on the oxide coating, but little material is deposited onto the channel walls for the coating growth. That is, the dielectric breakdown behavior has little contribution for coating thickening.

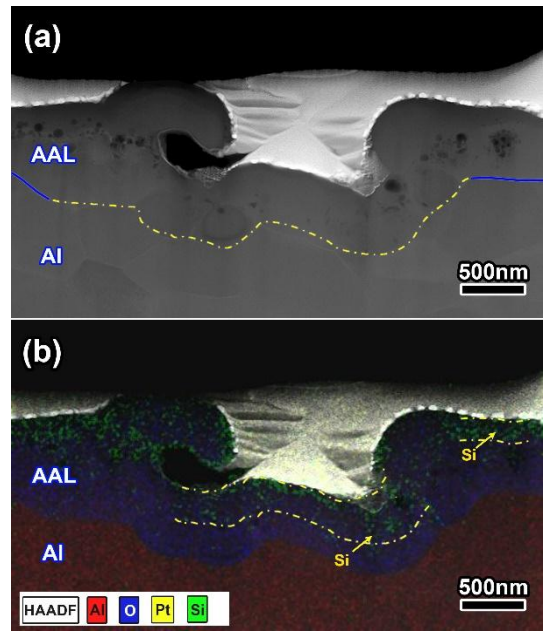

Fig. 6. (a) A cross-sectional HAADF-STEM micrograph and (b) the EDS element mapping of the breakdown pore in the PEO coating anodized for 12 s.

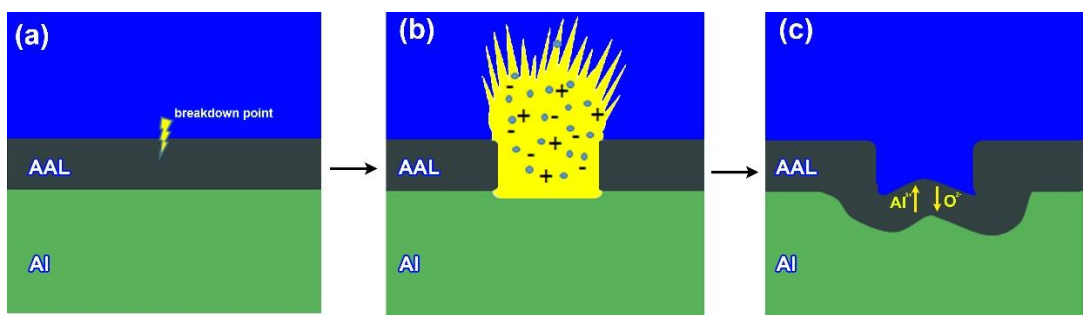

Fig. 7. Schematic illustration of the Ikonopisov's breakdown mechanism: (a) a breakdown occurs, (b) the electron avalanche causes a plasma evaporation and a loss of the local oxide coating, (c) the anodic film is rapidly formed at the bare spot.

3.4. Structure of the PEO coating in the stage II and the stage III.

After the BD-stage, the PEO process will enter the stage II. The most obvious characteristic of the stage II is the low increasing ratio of R-value and the continual surface discharge "flow" [18]. The surface discharge has been identified as the gas plasma by optical emission spectroscopy experiments [20, 23, 24]. The surface discharge flow moving randomly on the surface could create various morphologies on the coating structure. Fig. 8a is the SEM (BSE model) surface micrograph of the PEO coating anodized for 1 min (in the early stage II), showing two kinds of the surface roughness. The fine morphology marked by the number 1 and the rough morphology marked by the number 2 in Fig. 8a are enlarged in Fig. 8b and Fig. 8c, respectively. Meanwhile, two kinds of the surface morphologies are noticed in the sample of the PEO coating anodized for 5 mins (in the middle of stage II) as well, as shown by the numbers 3 and 4 in Fig. 8d and enlarged in Fig. 8e and 8f, respectively. Obviously, Figs. 8b, 8c and 8e show the similar surface morphology in series by an increasing surface roughness for the PEO coating in the stage II. However, Fig. 8f is the typical morphology of the "crater-like" pore, which means those areas have developed into the subsequent stage III at this moment [25].

To reveal the structure of the PEO coating in the stage II, the area numbered 2 in Fig. 8a was prepared into the cross-sectional TEM samples. The low magnification cross-sectional TEM image (Fig. 9a) shows that the coating contains a ~2000 nm thick AAL and an undulate surface layer with the thickness ranged in 50-300 nm. Figs. 9b-9d are the EDS element mappings of Al, O and Si, respectively. Both the AAL and the undulate surface layer mainly contain the elements of Al and O, while the Si element is only found at the surface layer and some big channels in the AAL. Fig. 9e is the enlarged morphology of the blue block in Fig. 9a with two numbered areas and the corresponding selected area electron diffraction (SAED) patterns (see the insets in Fig. 9e). It is obvious that the AAL is completely amorphous as shown by the inserted pattern from the number 2 area in Fig. 9e. Fig. 9f is the enlarged SAED pattern from the number 1 area in Fig. 9e showing the diffraction spots and rings superposed with an amorphous diffraction halo, in which all the polycrystalline diffraction rings are identified as the γ-$Al_2O_3$ phase. It indicates that the undulate surface layer is composed of the γ-$Al_2O_3$ nanoparticles and some amorphous $SiO_2$. Considering that the stage II does exist the surface plasma discharge flow [20], which would induce the outside part of the AAL to crystallize into the γ-$Al_2O_3$ nanoparticles, we name the undulate surface layer in the stage II as the plasma modification surface (PM-surface) layer. In the other hand, the discharge in stage II can be described as "discharge-on-surface" in brief, corresponding to the channel-free PM-surface layer. Fig. 9g is the enlarged micrograph of the red block in Fig. 9a, showing the typical ionic migration morphology of the PEO coating at the stage II. The corresponding AAL geometry of the PEO coating is shown in Fig. 9h. The distances between the pore bottom and the concave Al/AAL boundary is ~450 nm and the voltage is about 410 V (Fig. 2a), and the $E$ and $T$ in the AAL are estimated as ~0.91 V·$nm^{-1}$ and ~67 °C, respectively. These cross-sectional microstructure characterizations of the parallel-sided pores in the AAL and the concave boundary of Al/AAL reveal that the PEO coating growth in the stage II is also via the ionic migration mechanism inside the AAL, while the discharge-on-surface behavior leads to the crystallization of the AAL surface. Meanwhile the thin PM-surface layer indicates that the discharge-on-surface exists in a low energy and soft state [22].

It should be noticed that the micro-discharge in the PEO process is no longer designated the "breakdown" as in the traditional dielectric breakdown mechanism [1,13]. But here, when anodized for 1 min, the micro-discharge does exist in the PEO process and the coating does grow via ionic migration mechanism. It is obvious that the micro-discharge in the PEO process is not the presentation of dielectric breakdown, or rather the gas plasma, as revealed by optical emission spectroscopy experiments [20, 23, 24]. As revealed above, the growth of PEO coating in stage II is a process of combining discharge-on-surface with innermost ionic migration inside AAL.

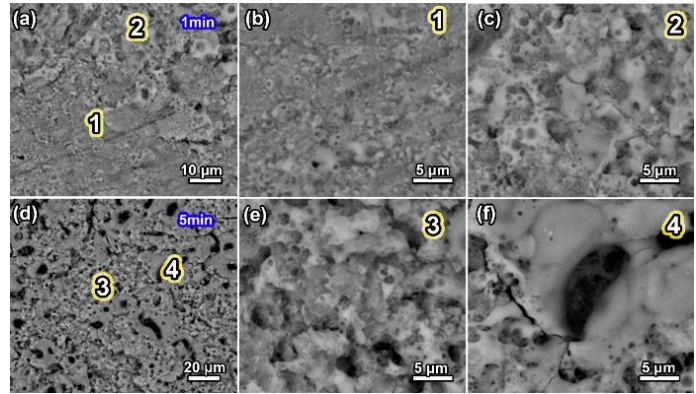

Fig. 8. SEM (BSE model) micrographs of the surface at the PEO coating. (a-c) anodized for 1 min, (b) and (c) are the enlarged areas in (a) numbered "1" and "2" respectively; (d-f) anodized for 5 min, (e) and (f) are the enlarged areas in (d) numbered "3" and "4" respectively.

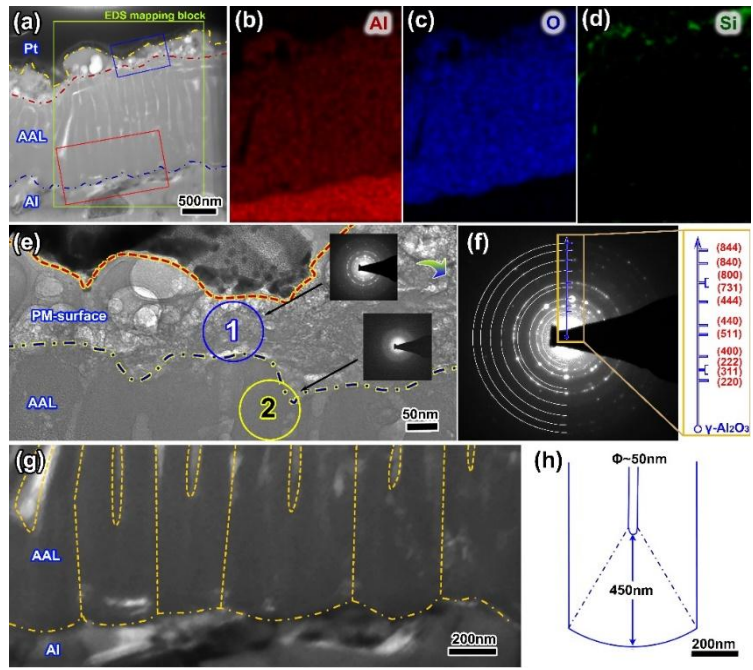

Fig. 9. (a) TEM cross-sectional micrograph of the PEO coating area numbered "2" in Fig. 8a and the corresponding EDS element mappings of (b) Al, (c) O and (d) Si. (e) The enlarged micrograph of the blue block in (a) with two insets of the SAED patterns corresponding to two annotated areas, (f) the enlarged SAED pattern of the area 1 in (e) showing the polycrystalline diffraction rings with the indices of $\gamma$-$Al_2O_3$ phase, (g) the enlarged micrograph of the red block in (a) and (h) the AAL geometry of the PEO coating.

Fig. 10a is the surface micrograph of the PEO coating anodized for 10 min, in which many "crater-like" pores on the coating reveals the PEO process evolved from the stage II to the stage III. When anodized for 20 min (Fig. 10b), the PEO coating developed into the stage III completely and formed the complete ceramic surface. The SEM cross-sectional micrograph of the PEO coating at 20 min (Fig. 8c) shows the typical structure of "out-layer, in-layer, AAL", and Figs. 8d-8f are the EDS element mappings of Al, O and Si, respectively. The TEM cross-sectional morphology, microstructure and composition of the PEO coating for 20 min (the stage III) have been comprehensively studied by using TEM in our previous work [25]. The formed PEO coating at the stage III is in turn composed of the inner AAL of $Al_2O_3$, the crystalline γ-$Al_2O_3$, and the bilayered aluminum silicate ($3Al_2O_3·2SiO_2$) with the nanometer-diameter channels at the in-layer and the micron-diameter channels at the out-layer [25]. The main driven force for the PEO coating growth in stage III was found to be the ionic migration behavior of the inner most AAL, combining with the localized micro-discharges at the out-layer channels of the aluminum silicate, which is called as the outside discharge-in-pore. The "crater-like" pores on the surface correspond to the micron-diameter channels in the out-layer at the stage III. These results imply that the evolution of the PEO coating from the stage II to III is a gradual process (Fig.8), as the "crater-like" pores initially appear at ~5 min and generate continuously until ~20 min. Therefore, it is always hard to give a clear boundary between the stage II and the stage III in the voltage vs. time curve (Fig. 1), although the microstructural characteristics at the two stages are clearly different.

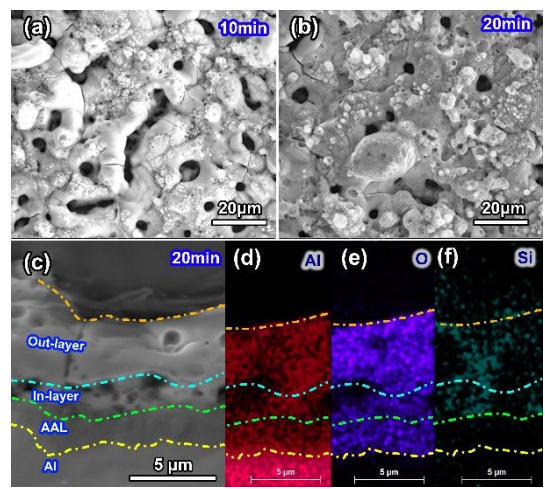

Fig. 10. SEM (BSE model) surface micrographs of the PEO coating anodized for (a) 10 min and (b) 20 min.

## 3.5. Formation and growth mechanisms of the PEO coating in various stages

According to structural characterizations at different stages as shown above, it is clear that the pathway for the PEO coating growth on aluminum is the ionic migration behavior of $Al^{3+}$ and $O^{2-}$ in the AAL driven by the high electric field strength in the three main stages I-III, except in the BD-stage. What makes these stages different is attributed to the difference in the role and influence of the surface discharge at the various stages.

In the stage I, the PEO coating, composed of the AAL alone, directly contacts with the alkaline electrolyte and uniformly grows in a short time (Fig. 11a). Under the strongly oxidizing (high $E$ and temperature) environment near the anode, the reactions at the AAL/solution interface occur as following [35]:

$$OH^-(aq) \rightarrow H_2O(aq) + O^{2-}(aq) \tag{1}$$

And the $O^{2-}$(aq) ions could enter the AAL (oxide) and join the ionic migration process:

$$O^{2-}(aq) \rightarrow O^{2-}(ox) \tag{2}$$

However, one should not consider the stage I in PEO as the conventional anodic oxidation process. As revealed in the section 3.2, the growth of the PEO coating in the stage I could cause an abrupt temperature rise in the coating, e.g. ~250 °C when anodized for 8 s at the current density of 0.1 mA·cm$^{-2}$. This is much different from a conventional anodic oxidation process in acid solution, in which the temperature rise from the ionic migration in oxides is only ~ 2 °C at a current density of 0.08 mA·cm$^{-2}$ [29, 36]. Moreover, the stage I in the PEO process is a transient process and lasts only 11 s in this case (as shown in Fig.2a), but the conventional anodic oxidation exhibits a steady growth process [29]. These differences are originated from the contrast of the interfacial electric field across the AAL/aqueous solution, which is determined by both the pH value of the aqueous solution and the surface isoelectric point (IEP) of the solid AAL.

Considering that the outermost surface of an oxide is covered with a layer of hydroxyl groups in aqueous solutions, these hydroxyl groups may remain un-dissociated when the pH value is the same as the surface IEP [37]. When the pH value of an aqueous solution is higher (or lower) than the IEP of oxide surface, the aqueous/solid oxide interfaces may form the negatively (or positively) charged surface with the following chemical reactions [38]:

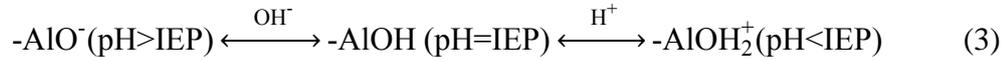

$$\text{-AlO}^-(\text{pH}>\text{IEP}) \xleftrightarrow{\text{OH}^-} \text{-AlOH (pH=IEP)} \xleftrightarrow{\text{H}^+} \text{-AlOH}_2^+(\text{pH}<\text{IEP}) \qquad (3)$$

Based on the reactions, the established surface electric field on the anodized coating has the opposing (same) direction with the applied electric field [39]. For the γ-Al$_2$O$_3$ with a surface IEP of ~7-8 [40], the conventional anodic oxidation performed in the acid solution (with a pH＜5 [41]) is significantly different from the PEO coating in the alkaline solution (always with a pH＞10). Therefore, the interfacial electric field in the stage I of the PEO process has an opposing direction with the applied electric field, which may act as an additional energy barrier to hinder the anions migration into the anode, and then results in the growth process in the stage I accompanied by the extra temperature rising and the luminescence. Consequently, the stage I is shortly terminated by the AAL breakdown (i.e. the BD-stage). The BD-stage (Fig. 7) is mainly dominated by the destruction of the AAL via dielectric breakdown and the formation of the breakdown-pits in the PEO coating.

In the stage II, the growth of the PEO coating is localized at the areas with the discharge-on-surface flow (Fig. 11b). The discharge could short-circuit the interfacial reactions of the AAL/alkaline solution and create the plasma influence on the outer layer of the coating, which results in the formation of an outmost PM-surface with the mixture of γ-Al$_2$O$_3$ nanoparticles and amorphous materials. Such composite surfaces often present a high activity, which has extended the PEO technology to the synthesis of new functional materials [9]. In addition, the PM-surface found in the stage II (Fig. 9e and Fig. 11b) strongly supports the assumption proposed in previous work [25], that the PEO process was combined with surface discharge and underneath ionic migration. That is, the surface discharge plasma cannot break down the amorphous Al$_2$O$_3$, but rather creates the high electric field strength for the ionic migration in the AAL.

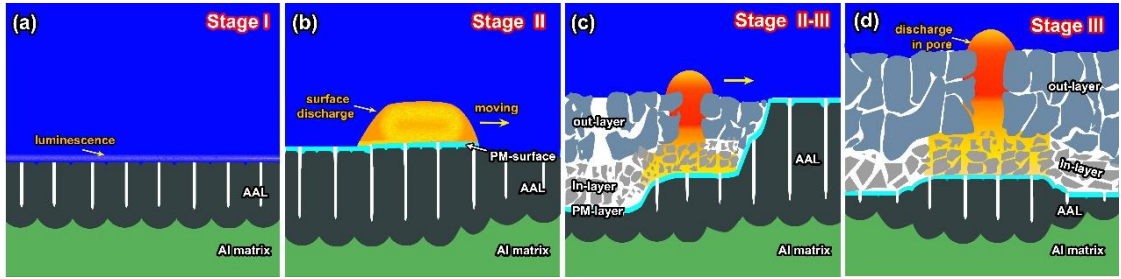

Fig. 11. Schematic diagrams illustrating the growth mechanism of the PEO coating in (a) Stage I with uniform growth, (b) Stage II with the growth localized under the moving surface discharges and the formation of the PM-surface, (c) Stage II-III with , and (d) Stage III.

From the stage II to the stage III, the cell voltage increases from ~400 V to 450 V (Fig. 1) and the surface discharge becomes more localized [18]. Also the oxygen emission in surface discharge plasma was found to be increased with the treatment time by using the optical emission spectroscopy [42]. This implies that the surface discharge is enhanced, and the thermal effect is getting stronger. Therefore, the flowing discharges at the end of stage II may cause the crystallization of amorphous $Al_2O_3$. In the meantime, the channels are created in the crystalline $Al_2O_3$ layer because of the volume shrinkage of at least 6.8% from amorphous alumina to crystalline $\gamma$-$Al_2O_3$ [43], and the graded structures of "out-layer, in-layer, AAL" are formed by the temperature gradient of the plasma heating in channels (shown in Fig.11c).

In the stage III, the discharges are further localized in the micron-sized pores [18, 25], which results in the formation of the ceramic coating containing Al and Si (element from the electrolyte solution) for enhancing the abrasion/corrosion resistance and the thermal barrier of materials (Fig. 11d). Despite the fact that the outside discharges generate enough heat to anneal the grain size of the out-layer into micron-size, the inner discharges at the inner-layer/AAL surface are still quite mild [22] and can only create a PM-layer with nanoparticles and amorphous materials on the AAL surface [25]. Hence, the stage III can be considered as the continuation of the stage II.

In general, the PEO coating growth on aluminum is via the ionic migration in the AAL driven by the high electric field strength in the three main stages I-III. The whole process of the PEO coating includes the uniform growth in the stage I, the localized destruction by the dielectric breakdown in the BD-stage, the localized growth induced by the discharge-on-surface flow in the stage II, and the more localized growth driven by the discharge-in-pore in the stage III. The transformation from the uniform growth in the transient stage I to the localized growth in the sustainable stage III reveals that the localized growth (or localized reaction) offers a chance for the main coating surface to dissipate heat into the electrolyte system, which is the key to achieve the sustainable PEO process at a moderate temperature range.

**Conclusion**

Based on the comprehensive SEM and TEM investigations, the formation mechanisms in the whole process of the PEO coating have been revealed as following. (i) In the stage I, the coating grows via the ionic migration, resulting in a ~1000 nm-thick AAL featured by the pore arrays of several tens nanometers in diameter; however, the stage I is unsustainable and will be shortly terminated by the AAL breakdown, which is different from a conventional anodic oxidation process in acid solutions. (ii) In the following breakdown stage, the coating is broken down and micro-size pits are formed in coating, which follows the Ikonopisov's breakdown mechanism. (iii) In the stage II, the coating grows via the discharge-on-surface flow and the inner ionic migration of the AAL, where a discharge influence layer composed of $\gamma$-$Al_2O_3$ nanoparticles and amorphous materials is formed at the exposed surface of the AAL coating. (iv) In the stage III, the coating grows steadily via the localized discharge-in-pore and the inner ionic migration of the underneath AAL, to form a gradient ceramic coating composed of the innermost AAL and the crystalline in-layer and out-layer with channels. The proposed mechanisms greatly enrich the knowledge about the early stages in the PEO process, uncover the importance of the ionic migration for the whole PEO process, and improve the understanding of the PEO coating growth by combining outside discharge with inside ionic migration. These results are also beneficial to the mechanism research and morphological control of the PEO coating on Mg, Ti and their alloys.

**Acknowledgements**

This work is supported by the National Natural Science Foundation of China (Grants Nos. 11374028 and U1330112) and the Foundation for Innovative Research Groups (Grant No. 51621003), the National Key Research and Development Program of China (Grant No. 2016YFB0700700), the Scientific Research Key Program of Beijing Municipal Commission of Education (KZ201310005002), and the Cheung Kong Scholars Programme of China.